# THE ROLE OF MASSIVE AGB STARS IN THE EARLY SOLAR SYSTEM COMPOSITION

Josep M. Trigo-Rodríguez [1,2], Domingo Aníbal García-Hernández [3], Maria Lugaro [4,5], Amanda I. Karakas [6], M. van Raai [4], Pedro García Lario [7], and Arturo Manchado [8].

[1] Institute of Space Sciences (CSIC), Campus UAB, Facultat de Ciències, Torre C-5, parells, 2ª planta, 08193 Bellaterra, Barcelona, Spain. E-mail: trigo@ieec.uab.es

[2] Institut Estudis Espacials de Catalunya (IEEC), Ed. Nexus, Gran Capità 2-4, 08034 Barcelona, Spain

[3] Instituto de Astrofísica de Canarias, La Laguna, E-38200, Tenerife, Spain

[4] Astronomical Institute, University of Utrecht, Utrecht, The Netherlands.

[5] Centre for Stellar and Planetary Astrophysics, Monash University, Victoria, Australia

[6] Research School of Astronomy & Astrophysics, Mt. Stromlo Observatory, Weston Creek ACT 2611, Australia

[7] Herschel Science Centre. European Space Astronomy Centre, European Space Agency. Apartado de correos 78, Villafranca del Castillo. E-28080 Madrid, Spain.

[8] Instituto de Astrofísica de Canarias, La Laguna, E-38200, Tenerife, Spain and Consejo Superior de Investigaciones Científicas

## ABSTRACT

**We demonstrate that a massive asymptotic giant branch (AGB) star is a good candidate as the main source of short-lived radionuclides in the early solar system. Recent identification of massive (4-8 $M_\odot$) AGB stars in the Galaxy, which are both lithium- and rubidium-rich, demonstrates that these stars experience proton captures at the base of the convective envelope (hot bottom burning), together with high-neutron density nucleosynthesis with $^{22}$Ne as a neutron source in the He shell and efficient dredge-up of the processed material. A model of a 6.5 $M_\odot$ star of solar metallicity can simultaneously match the abundances of $^{26}$Al, $^{41}$Ca, $^{60}$Fe, and $^{107}$Pd inferred to have been present in the solar nebula by using a dilution factor of 1 part of AGB material per 300 parts of original solar nebula material, and taking into account a time interval between injection of the short-lived nuclides and consolidation of the first meteorites equal to 0.53 Myr. Such a polluting source does not overproduce $^{53}$Mn, as supernova models do, and only marginally affects isotopic ratios of stable elements. It is usually argued that it is unlikely that the short-lived radionuclides in the early solar system came from an AGB star because these stars are rarely found in star forming regions, however, we think that further interdisciplinary studies are needed to address the fundamental problem of the birth of our solar system.**

## INTRODUCTION

The formation of the solar system started with the collapse of a molecular cloud to produce a protostar surrounded by a disk of gas and dust (Cameron 1962; Elmegreen 1985). The chemical composition of the solids initially present in this collapsing cloud likely reflected the products of stellar evolution and outflow that occurred during Galactic history (Carlson and Lugmair, 2000). The isotopic composition, however, may be strongly marked by the local environment where the solar system formed. In this context, the abundances of short-lived radionuclides (SLN, with half lives shorter than ~ 2 Myr), inferred to have been present in the early solar system (ESS), are a stringent constraint on the birth and early evolution of our solar system. This is because their relatively short half lives do not allow the observed abundances to be explained by continuous Galactic uniform production (i.e., galactic chemical evolution processes), which, in turn, implies that some type of nucleosynthetic event must have occurred very close in time and space to the forming Sun.

The identification of SLN incorporated into chondritic components is a current hot topic in meteorite studies. Although they decayed a long time ago, their daughter products are found in meteorite components. The identification of these daughter isotopes allows us to obtain reasonable data on the abundances of SLN incorporated into chondritic materials. For example, the abundance of $^{26}$Al and $^{60}$Fe in different mineral phases of meteoritic components provide clues on the contribution of these SLN to the primordial heating of planetesimals. Such contribution has shaped the further evolution of planetesimals as we know from thermochronometry (see e.g. Trieloff et al., 2003). These studies demonstrate that the primordial heating was mainly originated by the energy released by the decay of short-lived isotopes. In any case, additional research is needed to provide clues on the initial abundances of these radioactive isotopes and also on the accretion time scales of the parent asteroids of primitive meteorites. A quick accretion of these bodies will favor the role of short-lived isotopes in differentiation processes occurred in large asteroids.

In Table 1 we provide a list of the SLN and other radioactive isotopes of interest detected in the ESS. Most observed ratios are derived from the study of Calcium-Aluminum-rich inclusions (CAIs), which are generally considered to be the first objects that formed in the solar system. According to the X-wind model proposed by Shu et al. (2001), CAIs were produced in the hottest regions of the protoplanetary disk (the "reconnection rings") by the continuous heating produced by periodic flare activity from the young Sun, and at the same time enriched in SLN via spallation reactions induced by irradiation of solar energetic particles. Grossman (1972) had proposed much earlier that CAIs (and chondrules) formed by partial evaporation of material exhibiting CI chondrite (solar) composition with gas-melt exchange during flash heating events produced in the nebula. A nebular shock front as modeled by Boss & Durisen (2005) would heat CI-precursor materials to temperatures higher than 1,800 K during a few hours, explaining the observed CAI mineralogy. Alexander (2003), following the hypothesis of partial evaporation of CI dustballs, provided a uniform explanation of chondrule and CAI formation relatively far from the Sun (2-3 AU), in a less restrictive environment than the Shu et al. (2001) model. In this case, the dust precursor of CAIs would be subjected to efficient mixing with interstellar material, particularly if the solar nebula formed in a dense stellar environment. Indeed, there is evidence that many low-mass stars form in large clusters (> 700 members) together with high-mass stars (Lada and Lada 2003). This argument was used later by Hester et al. (2004) to suggest that the presence of SLN (especially $^{60}$Fe with a half life of only 1.5 Myr) is direct evidence that the solar system formed in such an environment, where the ejecta from a core-collapse supernova (SNII) quickly mixed with the material from which the meteorites formed. Connolly (2005) also proposed a similar argument, where he invoked mixing of early solar-system material with SNII material.

The first clue of the presence of $^{60}$Fe in the ESS was obtained from measured excesses of $^{60}$Ni in CAIs, up to an extraordinarily high $^{60}$Fe/$^{56}$Fe ratio of $(1.6\pm0.5)\times10^{-4}$ (Birck and Lugmair 1988). However, a more moderate initial ratio of $4\times10^{-6}$ was inferred from bulk samples (Birck and Lugmair 1988). More recent studies of troilite (FeS) grains contained in the Bishunpur and Krymka chondrites indicate a $^{60}$Fe/$^{56}$Fe=$(1.08\pm0.23)\times10^{-7}$ (Tachibana and Huss 2003). Evidence for the presence of the radionuclide $^{26}$Al in CAIs, with an early solar system $^{26}$Al/$^{27}$Al ratio of $5\times10^{-5}$ (the "canonical" value; e.g., MacPherson et al. 1995) is provided by excesses of $^{26}$Mg that are correlated with $^{27}$Al. Among those CAIs containing detectable $^{26}$Mg excesses attributable to $^{26}$Al decay, most have a $^{26}$Al/$^{27}$Al ratio of 3-$6\times10^{-5}$. More recent results (e.g., Young et al. 2005) indicate that some samples display a "supra-canonical" value of $^{26}$Al/$^{27}$Al~$6\times10^{-5}$. On the other hand, the vast majority of chondrules do not contain evidence for live $^{26}$Al during their formation. Hence, the amount of live $^{26}$Al apparently decreased to a very low level during the formation of chondrules, suggesting that chondrules formed later than CAIs, even if some chondrules showing $^{26}$Mg excesses ($^{26}$Al/$^{27}$Al~8-$9\times10^{-6}$) have been identified in three ordinary chondrites (Hutcheon and Hutchison 1989; Russell et al. 1996).

Spallation reactions induced by energetic particles originating from the early Sun (Shu et al. 1996, 2001; Gounelle et al. 2006), and by Galactic cosmic rays are the likely origin of $^{10}$Be, because this nucleus is not synthesized in stars. The isotope $^{36}$Cl, whose abundance appears to be coupled to that of $^{10}$Be in meteoritic materials, is also difficult to produce in stars (see discussion in Wasserburg et al. 2006). It is possible that a proportion of the observed abundances of some low atomic mass short-lived species, including $^{26}$Al, $^{41}$Ca, and $^{53}$Mn, came from this process. However, there are several difficulties with this scenario: heterogeneity in the abundances of the SLN would result from variation in the irradiation flux and the effect of shielding of CAI cores by mantles, but has not been observed in any study so far. Moreover, if $^{26}$Al was produced by spallation, it should be homogeneously distributed over a relatively small rocky reservoir (Duprat and Tatischeff 2007). Note also that data from hibonite grains indicates that the production of $^{10}$Be is decoupled from that of $^{41}$Ca and $^{26}$Al (Marhas et al. 2002), indicating that spallation probably did not produce these isotopes. Another difficulty is explaining how mm- to cm-sized CAIs could have remained close to a turbulent early Sun long enough to receive the required irradiation fluxes without falling into it. High atomic mass nuclei are not efficiently synthesized by spallation due to their high coulomb barrier, hence, the confirmed high abundance of $^{60}$Fe in the ESS necessarily calls for the contribution of a nearby stellar object.

The presence of $^{60}$Fe and other SLN in primitive meteorites has been used as indirect evidence that material from (at least) one nearby SNII polluted our forming solar system (e.g., Connolly 2005). This idea originated from the argument that a SNII shock-triggered the collapse of the presolar cloud (Cameron and Truran 1977). However, the SNII pollution scenario is very uncertain, with different stellar models giving $^{60}$Fe yields differing by up to one order of magnitude (Limongi & Chieffi 2006). Furthermore, a self-consistent solution has not been yet been found for the ESS concentrations of the various SLN because a too high abundance of $^{53}$Mn, which originates from very deep layers of the star, is produced when assuming a SNII origin for $^{26}$Al and $^{60}$Fe, which originate from layers further out. This main problem is usually addressed by imposing a condition that only material located above a specific "injection mass cut" can be incorporated in the proto-solar cloud (see Meyer 2005, for details and discussion). Wolf-Rayet stars - stars of masses higher than > 40 $M_{\odot}$ suffering strong stellar winds during their main sequence phase - can also produce $^{41}$Ca, $^{107}$Pd, and $^{26}$Al, but they do not make $^{60}$Fe because the neutron density is not high enough to activate neutron captures on $^{59}$Fe (Arnould et al. 2006).

Asymptotic giant branch (AGB) stars can also produce a variety of the SLN including $^{26}$Al, $^{41}$Ca, $^{60}$Fe, and $^{107}$Pd. Low-mass (~1−3 $M_{\odot}$) AGB stars, however, cannot explain the $^{26}$Al ESS abundance unless some kind of extra mixing above that found from stellar models (also called cool bottom processing) is invoked (see discussion in Wasserburg et al. 2006). The origin of $^{26}$Al can be attributed to a massive (~6 $M_{\odot}$) AGB star (Lee et al. 1977; Nørgaard 1980) experiencing proton-capture nucleosynthesis at the base of the convective envelope (hot bottom burning, or HBB). AGB stars are low and intermediate-mass ($0.8 < M < 8$ $M_{\odot}$) stars in their final nuclear burning phase of evolution, and are located in the low temperature and high luminosity region of the Herzsprung-Russell diagram (see Iben and Renzini 1983; Herwig 2005 for reviews). After core helium burning, a low- to intermediate-mass star has transformed all helium (He) in the core into carbon and oxygen, the core contracts and the outer layers expand: the star is now on the AGB. Helium is ignited in a thin shell surrounding the C-O core, and, together with H-shell burning, provides most of the surface luminosity. During the AGB phase recurrent thermal instabilities, or thermal pulses (TP), develop in the thin He-burning shell and drive convection over the whole He-rich region between the H- and the He-burning shells (He intershell). Most of the energy produced by the pulse drives an expansion of the whole star, which can result in the convective envelope moving inwards (in mass) into the He-burnt region. This mixing episode is known as the third dredge-up (TDU) and enriches the stellar surface in the products of partial He burning, including $^{12}$C and $^{22}$Ne. If enough carbon is mixed to the stellar surface the star is transformed from O-rich to C-rich, where the C/O ratio > 1, and indeed this is the case for low-mass (~1−3 $M_{\odot}$) AGB stars. In more massive stars (> 4−5 $M_{\odot}$ depending on Z), HBB causes the star to retain an O-rich atmospheric composition. HBB models (Mazzitelli et al. 1999; Lattanzio et al. 1997; Karakas and Lattanzio 2003) also predict the production of $^{26}$Al and $^{7}$Li, low values for the $^{12}$C/$^{13}$C ratio (~3−4), the almost complete destruction of $^{18}$O, and excesses in $^{17}$O (Forestini and Charbonnel 1997). The AGB phase is terminated when extreme mass loss removes the H-rich envelope, at a few times $10^{-5}$ $M_{\odot}$ per year (Vassiliadis and Wood 1993). The end the evolution of these stars is represented by the post-AGB and planetary nebulae phases, followed by the stellar cores eventually cooling to become C-O white dwarfs.

*Slow* neutron capture nucleosynthesis (the *s* process) can also occur in the He-shells of AGB stars, allowing the synthesis of elements heavier than Fe. Thermally pulsing AGB star models are able to account for the cosmic origin of roughly half of all elements heavier than iron, with the models supported by observations of AGB stars showing enrichments of *s*-process elements such as Sr, Tc, Ba, and La (Busso et al. 2001). Two main reactions, $^{13}$C(α,n)$^{16}$O and $^{22}$Ne(α,n)$^{25}$Mg, provide free neutrons in the region located between the H and the He burning shells (the He intershell). The $^{22}$Ne neutron source requires higher temperatures (> 300 million degrees) and provides higher neutron densities (up to $10^{13}$ n/cm$^{3}$) than the $^{13}$C source; on the other hand, the $^{13}$C neutron source provides a total number of neutrons higher than the $^{22}$Ne source. While the $^{22}$Ne source is likely to be activated in massive AGB stars, the $^{13}$C source is inferred to produce the bulk of the *s*-process elements in low-mass AGB stars (see Lugaro and van Raai 2008 for a recent review). The high-neutron density coming from the $^{22}$Ne neutron source in massive AGB stars is capable of activating branchings on the *s*-process path that result in the production of neutron-rich isotopes such as $^{60}$Fe, $^{86}$Kr, $^{87}$Rb, and $^{96}$Zr. The abundance of Rb relative to other nearby *s*-process elements (e.g., the Rb/Zr or Rb/Sr ratio) is sensitive to the neutron density and as such, represents a discriminant for the operation of the $^{22}$Ne versus the $^{13}$C neutron source. The low Rb abundances seen in the majority of *s*-process rich AGB stars has been thus used to conclude that the $^{13}C(\alpha,n)^{16}O$ reaction is the main neutron source for the *s* process, and that these stars have low initial masses, which is also supported by their luminosities (Lambert et al. 1995; Abia et al. 2001).

There is now observational evidence that Rb is highly enriched in massive (~4-8 $M_{\odot}$) AGB stars, likely due to the production of $^{87}$Rb (García-Hernández et al. 2006). Using these results, García-Hernández et al. (2006) concluded that a massive AGB star in the vicinity of the early solar system could have induced fluctuations in the Rb/Sr ratio of primitive chondritic materials. In the present paper we use this recent observational confirmation that the $^{22}$Ne neutron source and TDU are occurring in massive AGB stars to explore the possible role of these stars in the composition of the ESS. We discuss the production of SLN from a massive AGB star of 6.5 $M_{\odot}$, as well as the possible effects of such pollution on stable isotopic ratios. Given the often-discussed issue of the implausibility of an AGB star being near a forming solar-type star, we review the literature and add some discussion on this point in the appendix, where we also summarize the current studies on supernova pollution.

Finally, we discuss the role of massive AGB stars as a potential site for the origin of some presolar grains recovered from primitive chondrites showing extremely anomalous composition, with respect to the bulk of the solar system material (see, e.g., Clayton and Nittler 2004 for a review). These grains condensed in stellar outflows and explosions as the gas cooled, and were part of the solar nebula material before the consolidation of the first meteorites.

## METHODS AND MODELS TO STUDY MASSIVE AGB STARS

### Precedents

García-Hernández et al. (2006, 2007a) recently determined the Li, Zr, and Rb abundances in a large sample of massive Galactic O-rich AGB stars belonging to the class of OH/IR stars (i.e., stars extremely bright in the infrared showing OH maser emission) from stellar spectra obtained using a high-resolution optical spectrograph. By fitting the spectra these authors derived the fundamental parameters of these stars (e.g., the effective temperature $T_{eff}$~2700–3300 K and the metallicity Z~0.02) as well as their nucleosynthesis pattern. The estimated Rb/Fe ratios (up to 100 times larger than solar) provide the opportunity to study the influence of these stars on the chemical enrichment of the interstellar medium, as well as to test theoretical models. The observed correlation (see Figure 2 of García-Hernández et al. 2006) between the Rb abundances and the OH expansion velocities, which can be taken as a distance- independent mass indicator, confirms that the efficiency of the $^{22}$Ne neutron source is directly correlated with the stellar mass, as predicted by our massive AGB nucleosynthesis models (van Raai et al. 2008). However, the largest Rb enhancements observed in some stars are not matched by our present solar metallicity models. These stars may represent a stellar population of even higher mass.

In addition, the observed Rb overabundances are coupled with only mild excesses of Zr ([Zr/Fe][1]<0.5 dex) in these massive AGB stars. This is an important observational constraint for our theoretical AGB model and indicates that the efficiency of the $^{13}C$ neutron source is extremely low in these stars (see the following sections for more details). Note that this is in contrast to the lower-mass AGB stars, such as the S-type AGB (C/O~0.7-0.95) stars, which show strong Zr overabundances ([Zr/Fe]> 1.0 dex) and where $^{13}C$ is the dominant neutron source at the s-process site.

Recent models of massive AGB stars

For this study we use a massive AGB stellar model of 6.5 $M_{\odot}$ with a solar (Z = 0.02) initial composition, chosen because it is the most massive out of the Z = 0.02 models computed by Karakas and Lattanzio (2007), and thus has the shortest lifetime (~54 Myr), and because it has HBB and efficient TDU mixing. The stellar structure was calculated with the Monash version of the Mount Stromlo Stellar Structure Program where the numerical method used to compute the stellar models has previously been described in detail (Karakas et al. 2006; Karakas and Lattanzio 2007). Important model parameters include the treatment of convection and the mass-loss rate. The mass-loss rate determines the AGB lifetime along with the duration of HBB, and on the AGB we used the observationally-based formulation provided by Vassiliadis and Wood (1993), which results in fairly low outflow rates ($10^{-7}$ $M_{\odot}$ per year) until the start of a "superwind" phase, when the mass-loss rate increases to a few times $10^{-5}$ $M_{\odot}$ per year.

The structure in convective regions is determined using the mixing-length theory which depends on the parameter $\alpha$ - the mixing length divided by pressure-scale height, set at 1.75 in our models - with the assumption of instantaneous mixing. The structure of the convective envelope has been shown to be sensitive to the choice of convective model, along with the choice of $\alpha$, with larger values of $\alpha$ resulting in more efficient convection (Ventura and D'Antona 2005). The amount of TDU mixing in AGB models is dependent on the numerical treatment of the border between the radiatively stable He-intershell and the convective envelope following a TP (Frost and Lattanzio 1996; Mowlavi 1999). The amount of mixing taking place between the H-exhausted core and the envelope is defined by the TDU efficiency, which is the ratio between the amount of matter dredged into the envelope divided by the amount by which the H-exhausted core grew during the previous interpulse period; efficient dredge-up has this ratio close to unity. That the TDU does occur is well supported by observations of C-rich AGB stars, and stellar evolution codes that do not include some mixing beyond the inner boundary of the convective envelope defined by the Schwarzschild criterion may not see the TDU. Formally, the Schwarzschild boundary is located where the adiabatic and radiative temperature gradients are equal. In stellar models, however, a discontinuity develops (see Frost and Lattanzio 1996 for details) which inhibits the inward movement of the envelope and the ability to find the point closest to neutral buoyancy, which is where the ratio of the radiative to adiabatic temperature gradients are equal to unity. Lattanzio (1986) implemented a technique to search for this neutral border in our code and this has since been shown to increase the efficiency of the TDU compared to models that strictly use the Schwarzschild criterion (Frost and Lattanzio 1996).

The 6.5 $M_{\odot}$ model was computed from the zero age main sequence to near the tip of the AGB. We assumed initial solar abundances from the compilation of Anders and Grevesse (1989). During the AGB, the model experienced 39 TPs and 36 episodes of efficient TDU mixing, where the efficiency parameter was above 0.8 for 29 TPs (see tabulated data in Karakas and Lattanzio 2007). Even with such efficient mixing, the 6.5 $M_{\odot}$ model mixes about a factor of two less core matter into the envelope than a 3 $M_{\odot}$ model of the same composition, owing to the much smaller mass of the He-intershell region mass (typically a factor of 10 smaller). The 6.5 $M_{\odot}$ model experienced HBB with a maximum temperature at the base of the envelope of 86 million K. This temperature ensures that the $^{12}C$ mixed to the envelope via TDU is converted into $^{13}C$ and $^{14}N$, preventing the formation of a C-rich atmosphere.

---

[1] $[X/Y]=\log(X/X_{\odot})-\log(Y/Y_{\odot})$.

With the 6.5 $M_{\odot}$ structure as input, we computed several nucleosynthesis models in order to obtain the evolution of the most relevant SLN. In the post-processing nucleosynthesis code we use a time-dependent convective mixing model in convective regions, with the location of boundaries between convective and radiative regions provided by the stellar structure code. We employed two different networks in order to minimize the computational time. One network includes 207 species, from protons to sulfur and from the iron peak to palladium, and 1650 reactions. This was needed to specifically evaluate the $^{107}Pd$ abundances. The other network includes 125 species, from protons to the iron peak, and 1000 reactions, and was used to evaluate the $^{41}Ca$ abundance. From both networks we obtain the abundances of $^{60}Fe$ and $^{26}Al$. We make use of neutron sinks to account for the missing species in each network, although, as discussed by Karakas et al. (2007), the choice of neutron sinks and their neutron capture cross sections do not significantly affect the final results. We have made a new update of our reaction library: starting from the library described in Karakas et al. (2007) we have further included neutron capture cross sections from the Bao et al. (2000) compilation as well as the latest $^{41}Ca(n,\alpha)^{38}Ar$ rate (de Smet et al. 2006), by far the main channel for the destruction of $^{41}Ca$ in neutron-rich conditions, and the latest $^{36}Cl(n,p)^{36}S$ and $^{36}Cl(n,\alpha)^{33}P$ rates (de Smet et al. 2007). For electron captures on $^{41}Ca$ we took the terrestrial mean half-life of 0.14 Myr. In fully ionized stellar conditions the decay time is longer, unless the density is higher than a few $10^4$ gr/cm$^3$ (Fuller et al. 1982), which is of the order of the maximum value reached in the He intershell of massive AGB stars. Hence, it should be kept in mind that our estimated $^{41}Ca$ abundance is a first-order approximation and we plan to implement a more accurate description of the decay rate of $^{41}Ca$ in future calculations.

## RESULTS AND DISCUSSION

## MODEL PREDICTIONS AND COMPARISON TO ESS CONSTRAINTS

### Model comparison to SLN in the ESS

We have processed the isotopic yields from our massive AGB nucleosynthesis model following the same procedure described in detail by Wasserburg et al. (2006). As discussed at length by these authors, we can restrict our first analysis to four ratios involving radioactive nuclei in the ESS: $^{26}Al/^{27}Al$, $^{41}Ca/^{40}Ca$, $^{60}Fe/^{56}Fe$, and $^{107}Pd/^{108}Pd$[2]. In Table 2 we give the abundance ratio, $(N^R/N^I)_{ENV}$, of each radioactive isotope R to the chosen stable isotope I in the stellar envelope at the end of the computed AGB evolution. The production factors of the stable isotopes, $q^I_{ENV}/q^I_0$, which are the ratio of the final surface abundance to the initial (solar) abundance are within 2% of unity, except for $^{27}Al$ that is overproduced by 8%. Our model confirms that the abundances of $^{129}$I and $^{182}$Hf, as well as of the lighter $^{53}$Mn, are not produced in AGB stars. Hence, they can only be explained by Galactic uniform production in the AGB pollution scenario. The presence of $^{10}$Be requires irradiation in the ESS, and this may be the same for $^{36}$Cl, which does not accompany $^{26}$Al in meteoritic materials.

[2] We treat $^{107}Pd$ together with the SLN nuclei, even though its half live is 6.5 Myr.

In a similar procedure to Wasserburg et al. (2006), we consider a model without the inclusion of a $^{13}$C pocket. A $^{13}$C pocket is a tiny region in the upper layers of the He-intershell where the $^{13}$C($\alpha$,n)$^{16}$O reaction is assumed to operate. Note that the amount of $^{13}$C left over by CN cycling is not enough to activate an efficient *s* process (Gallino et al. 1998). Thus, it is hypothesized that some partial mixing of protons from the envelope penetrates the He intershell at the end of each TDU episode, when a sharp discontinuity arises between the convective envelope and the radiative intershell. This extra mixing leads to the formation of $^{13}$C via proton captures on the abundant $^{12}$C. The mixing mechanism responsible for producing the pocket is unknown although rotation, convective overshoot and gravity waves have all been suggested (Busso et al. 1999; Herwig 2005). Wasserburg et al. (2006) demonstrated that if a $^{13}$C pocket is included the Pd yield increases so greatly that it is not possible to find a simultaneous solution for the ESS abundances of $^{107}$Pd and of the other lighter nuclides. The choice of not including a $^{13}$C pocket in our massive AGB model is also supported by observations of Zr in massive galactic O-rich AGB stars (Garcia-Hernandez et al. 2007a). The [Zr/Fe] ratios are found to be solar within 0.5 dex, and our model can only match this constraint if the $^{13}$C pocket is not included.

In Figure 1 we illustrate the evolution of the isotopic ratios of interest. $^{26}$Al is already increased at the stellar surface during the second dredge-up. This mixing episode occurs after core He burning and carries the $^{26}$Al that was produced by H burning from regions deep in the star, just above the core, into the envelope. The abundance of $^{26}$Al is further enhanced by HBB during the AGB phase. $^{41}$Ca, $^{60}$Fe, and $^{107}$Pd are instead produced by neutron captures. $^{107}$Pd, and $^{41}$Ca to a smaller extent, are already increased when the second dredge-up carries to the surface material that experienced a small neutron flux due to the activation of the $^{13}$C($\alpha$,n)$^{16}$O reaction in the deep layers of the star during core He burning. During the second dredge-up the envelope of the 6.5 $M_\odot$ model penetrated to a depth of 0.95 $M_\odot$, at which point in mass the temperature reaches ~$10^8$ K thus activating the $^{13}$C($\alpha$,n)$^{16}$O reaction on the $^{13}$C left over by previous H burning. The $^{107}$Pd and $^{41}$Ca abundances are further enhanced during the AGB phase, together with that of $^{60}$Fe, when neutrons are released during TPs by the $^{22}$Ne neutron source, and the TDU carries material from the intershell into the envelope. Note that $^{60}$Fe can only be produced in the TPs because the high neutron densities of up to $10^{13}$ n/cm$^3$ allow the efficient activation of the branching point at $^{59}$Fe. Our results (Table 2) are different from the solar metallicity 5 $M_\odot$ model of Wasserburg et al. (2006) because of the following:

- The initial stellar mass is different;
- The $^{26}$Al/$^{27}$Al ratio is about 30 times higher in our model because HBB is at work;
- The $^{41}$Ca/$^{40}$Ca is roughly 30% higher. This may be due to our inclusion of the recent estimate of the $^{41}$Ca(n,$\alpha$)$^{38}$Ar reaction rate, which is rougly 30% lower than previous estimates;
- The $^{60}$Fe/$^{56}$Fe and $^{107}$Pd/$^{108}$Pd ratios are roughly four times smaller, and also the overproduction factor of $^{107}$Pd is 60% lower. This is because of two different effects. First, our $^{22}$Ne($\alpha$,n)$^{25}$Mg reaction rate is from Karakas et al. (2006), which is lower than previous estimates and this accounts for more than a factor of two difference;

- Second, our evolution stops at 39 TPs owing to the adoption of the stronger mass loss from Vassiliadis and Wood (1993), in agreement with the recent observations of strongly obscured OH/IR massive AGB stars. This is compared to 48 TPs for the model of Wasserburg et al. (2006), where the mass-loss rate from Reimers (1975) was used.

Considering the numbers from Table 2 and following the same procedure as Wasserburg et al. (2006) we obtain from $^{107}$Pd an allowed range for the dilution factor of $1.65\times10^{-3} < f_0 < 3.48\times10^{-3}$, where $f_0=M_{ENV}/M_0$ and $M_{ENV}$ is the mass of injected stellar envelope, and $M_0$ is the mass of the cloud. We chose a value of $f_0=3.3\times10^{-3}$ because this will provide us with the best fit to the observations. This value corresponds to a dilution factor of 300, which is close to the dilution factor of 100 found in the hydrodynamics models of Boss (1995) where an AGB star triggers the formation of the solar system while injecting into it 0.01 $M_\odot$ of material. More precisely, in our case, the AGB star would inject less than 1% of its envelope into a presolar cloud of 1 $M_\odot$. Then, we employ $^{41}$Ca to derive $\Delta_1$, which is the time interval between injection of the radionuclides and formation of the first solid bodies in the solar system, to obtain a value of 0.53 Myr. There are no direct indications for the value of $\Delta_1$, however, as discussed by Wasserburg et al. (1995), a free-fall time of the order of 0.5 Myr corresponds to densities of the order of 8000 H atoms cm$^{-3}$, which are within the range observed in dense molecular clouds. Finally we determine $\Delta_2$, which is the time interval between the initial state and the time of formation of differentiated objects, to be equal to 6.0 Myr assuming that at such time $^{107}$Pd/$^{108}$Pd $= 2 \times 10^{-5}$, as observed. Also, it is not possible to obtain direct indications for the value of $\Delta_2$ because the lifetimes of long-lived radioactive nuclides, which can be used as clocks, are not known to the precision required to obtain an accuracy of ~1 Myr (see Wasserburg et al. 2006 for discussion).

With the chosen values for $f_0$, $\Delta_1$, and $\Delta_2$ we find a self-consistent fit to the observed ratios in the ESS for the four radionuclides considered here, as shown in Table 3. We note that Wasserburg et al. (2006) had to assume the $^{26}$Al/$^{27}$Al ratio in the AGB envelope since their models do not produce $^{26}$Al in the needed amount, while with our model we obtained a self-consistent solution also for this isotope. The difference of 56% between our $^{26}$Al/$^{27}$Al ratio at $\Delta_1$ and the observed value is well within the model uncertainties. Considering only nuclear uncertainties, the errors bar of the $^{25}$Mg(p,$\gamma$)$^{26}$Al reaction rate produces an uncertainty of approximately 50% in the yields of $^{26}$Al, for similar models to the 6.5 $M_\odot$ model presented here (Izzard et al. 2007). The $^{26}$Al(p,$\gamma$)$^{27}$Si reaction rate is even more uncertain, with a typical error bar of three orders of magnitude in the temperature range of HBB. Using the upper limit suppresses $^{26}$Al production by HBB by two orders of magnitude (Izzard et al. 2007). Moreover, the treatment of convection during HBB affects the nucleosynthetic results. In particular, using the "full spectrum of turbulence" model instead of the classic mixing-length theory to describe the AGB envelope convection results in higher HBB temperatures leading to a higher production of $^{26}$Al (see discussion in Ventura and D'Antona 2005). For the other isotopes, the main nuclear uncertainties come from the neutron capture reaction rates, in particular for the unstable isotopes there are only theoretical estimates available (with the notable exceptions of $^{41}$Ca and $^{36}$Cl reported in the previous section), while the main stellar uncertainties are related to the efficiency of the TDU and the mass-loss rate.

Finally, we note that our model produces a final $^{36}Cl/^{35}Cl \sim 10^{-4}$ at the stellar surface, which is too low by a factor of ~50 to explain the ratio observed in the solar system with the same values for the dilution and the time intervals used above. This isotope is also problematic for the SNII pollution scenario (Meyer 2005). Uncertainties in the neutron capture cross sections, which may play a role in the stellar prediction for the abundance of this nucleus, need to be carefully analyzed, as well as its possible production via spallation in the ESS.

We conclude that a massive AGB star is a good candidate for having polluted the ESS with radioactive nuclei, and that further investigation is required. In a forthcoming paper we will discuss results for models of different masses and metallicities than the model presented here, examine stellar and nuclear uncertainties in more detail (for $^{26}$Al and $^{60}$Fe more models and discussion can be found in Lugaro and Karakas 2008), and discuss other heavy radioactive nuclei of interest such as $^{135}$Cs and $^{205}$Pb, which are produced by AGB stars.

Model predictions for stable isotope anomalies and other radioactive nuclei

If the ejecta of a massive AGB star polluted the early solar system, not only the abundances of radioactive isotopes would have been altered, but also those of stable isotopes. Such anomalies would be much less evident because of the large dilution; however, it is of interest to spot correlations from which it may be feasible to discriminate among the different pollution scenarios. In Table 4 we present predictions for anomalies for all stable isotopic ratios included in our network, in form of variations with respect to solar in parts per ten thousand (ε), a unit widely used when measuring very small anomalies with respect to solar, together with their final computed ratios ($Y^i_{AGB}/Y^j_{AGB}$) at the surface of the 6.5 $M_\odot$ model, the AGB ratios diluted with solar system material by 1/300 (the dilution factor derived in the previous section), and the solar ratios ($Y^i_\odot/Y^j_\odot$) that we have used as references and as initial values in our calculations (Anders and Grevesse 1989).

Overall the anomalies (ε) are very small, within 2.4%, and typically smaller than those expected from a scenario where a SNII polluted the protoplanetary disk: Gounelle and Meibom (2007) derived O isotopic anomalies from SNII pollution varying from 1% up to 22%, depending on the details of the scenario employed. In the AGB case the largest anomalies are associated with the C, N, and O isotopic ratios. These show the effect of HBB in that $^{13}$C, $^{14}$N, and $^{17}$O are enhanced, while $^{18}$O is depleted. The Ne and Mg isotopic anomalies represent the combined effect of TDU and HBB, while all the remaining ratios are altered by neutron captures driven by the $^{22}$Ne source in the TP combined with the TDU. Typically, these result in excesses in the neutron-rich isotopes produced during the *s* process, for example in the case of $^{46}$Ca, $^{58}$Fe, the Ni isotopes, $^{86}$Kr, $^{87}$Rb, and $^{96}$Zr, and deficits in isotopes attributed to the proton-capture process (*p* process), as in the cases of $^{74}$Se and $^{78}$Kr, or to the *rapid* neutron-capture process (*r* process), as in the case of $^{100}$Mo. For stable nuclei produced by the decay of SLN, we have also calculated the anomaly obtained by adding the abundance of the radioactive isotope to the stable isotope. In the case of $^{41}$K and $^{60}$Ni this increases the anomaly by a few parts per ten thousand, and in the case of $^{99}$Ru this makes a very large difference, turning the anomaly from positive to negative. In Table 5 we present predictions for other SLN of interest, similarly to what presented in Table 3. The model predicts large excesses of $^{93}$Zr and $^{99}$Tc, which are located on the main *s*-process path.

The predicted positive correlations between $^{60}$Ni and $^{62}$Ni can be compared to CAIs data. Quitté et al. (2007) noted a correlation between these two isotopes with slope 0.53, while our model predicts 0.13, in the case where we just consider the abundance of $^{60}$Ni, or 0.34, in the case when add the abundance of $^{60}$Fe to that of $^{60}$Ni. However, it is not clear if $^{60}$Fe should be taken into account when making such comparisons because the Fe/Ni ratio is not constant in the measured CAIs. Moreover, neutron capture cross-sections in the Fe, Ni region have significant uncertainties (Bao et al. 2000), which need to be tested. We also predict a positive correlation between $^{62}$Ni and $^{96}$Zr with a slope of ~0.8. Such a correlation may also be present in CAIs (Quitté et al. 2007), although the error bars are too large for a positive identification.

Also of interest are the Rb and Sr isotopic anomalies. This is because among different CAIs small variations of up to 3ε units in the inferred initial $^{87}$Sr/$^{86}$Sr ratios have been measured (Podosek et al. 1991), and these may be qualitatively explained by heterogeneity due to pollution of massive AGB material. Our predicted negative values for the Sr isotopic anomalies from the 6.5 $M_{\odot}$ model are in qualitative agreement with the results of Lugaro et al. (2003). From Table 4, we expect a relatively large $^{87}$Rb/$^{86}$Sr anomaly, however, given the time intervals between $\Delta_1$ and $\Delta_2$ considered here, we do not expect any radiogenic contribution from $^{87}$Rb to $^{87}$Sr. We do predict variations up to 6ε in the $^{87}$Sr/$^{86}$Sr ratio itself, although this should be carefully tested against model and nuclear uncertainties. Other explanations for the observed variations, such as elemental Rb/Sr fractionation, are also possible.

Model comparison to stellar grain evidence

A detailed discussion of the stellar grain evidence contained in primitive meteorites is also required because they retain the isotopic signatures of their parent stars and thus provide constraints on the models of nucleosynthesis that we are introducing here. For example, it is well known that most presolar SiC grains have isotopic anomalies that suggest their formation in ~1.5 to 3 $M_{\odot}$ AGB stars (see e.g. for a review Zinner, 2003), but it remains unexplained why we have not been able to clearly identify stellar grains from AGB stars with masses over 4 $M_{\odot}$. Many questions are open in reference to this topic, but perhaps the peculiar chemical composition of these intermediate mass stars, or some unknown process prevents the survival of stellar grains that form around these objects.

As outlined above, massive O-rich AGB stars pertain to the group of OH/IR stars, which are considered to be the second most important source of dust in the Galaxy after WR-type stars (see, e.g., Alexander 1997). However, there is still no conclusive evidence that any stellar grains recovered so far from primitive meteorites originated in one of these stars. Most presolar stellar grains show the signature of originating in low-mass red giant stars and AGB stars. For carbonaceous grains, such as silicon carbide (SiC) grains, it is not easy to associate them with massive AGB stars because HBB prevents the formation of a C-rich atmosphere, which is the necessary condition for SiC to form. Thus, the nucleosynthesis pattern of massive AGB stars may be found in presolar oxide grains (Nittler et al. 1994, 1997; Lugaro et al. 2007), even though the possibility of forming C-rich grains in O-rich environments is still an open issue. Recent detailed dynamical models indicate that the formation of C-rich grains in the envelopes of O-rich AGB stars cannot be completely discarded due to non-equilibrium effects (Höfner and Andersen 2007). In this case, the isotopic signature of some A+B SiC grains (Amari et al. 2001) showing low values of the $^{12}$C/$^{13}$C ratios (<10), high $^{26}$Al/$^{27}$Al ratios in the range ~$10^{-3}$ – $10^{-2}$, and high $^{14}N/^{15}N$ ratios (up to $10^4$), might have been produced by HBB.

For the relatively massive AGB stars that do not show the effects of HBB (~4-5M$_\odot$), comparison of the isotopic composition of Sr, Zr, Mo, and Ba in single mainstream SiC grains with theoretical *s*-process predictions excludes these stars as the parent stars of mainstream SiC grains (Lugaro et al. 2003). For example, as described above, $^{96}$Zr is overproduced in the envelopes of these stars with respect to the other Zr isotopes and with respect to solar, while measured single SiC grains all show deficits in this neutron-rich isotope.

The relatively recent development of a new type of ion probe, the NanoSIMS, has led to the recovery of sub-μm presolar oxide grains (Zinner et al. 2003, 2005). Among them, one spinel grain (OC2) might have been produced in a massive AGB star experiencing HBB (Zinner et al. 2005). This hypothesis can explain the Mg and Al composition of this peculiar oxide grain, which shows enhancements in the $^{25}$Mg and $^{26}$Mg isotopes compared to solar coupled with extreme $^{17}O/^{16}O$ and $^{18}O/^{16}O$ ratios, as expected from the combined activation of the $^{22}Ne(\alpha,n)^{25}Mg$ and $^{22}Ne(\alpha,\gamma)^{25}Mg$ reactions (Karakas et al. 2006) and HBB (Forestini and Charbonnel 1997). The composition of OC2 was quantitatively matched using a massive AGB model (~6 M$_\odot$) within the nuclear error bars associated to the $^{16}$O+p reaction rate (Lugaro et al. 2007). However, a new evaluation of these error bars appears to exclude a massive AGB origin for the grain OC2 (Iliadis et al. 2008).

## SUMMARY AND CONCLUSIONS

The recent identification of Rb-rich ~4-8 M$_\odot$ AGB stars has provided observational evidence that the $^{22}Ne(\alpha,n)^{25}Mg$ reaction is indeed the dominant neutron source in massive AGB stars, and the third dredge-up is activated together with hot bottom burning. Based on this finding, we suggest that a massive (~6.5 M$_\odot$) AGB star, with efficient neutron-capture processing, and experiencing both previously mentioned processes, might have been the source of short live nuclides in the ESS.

Summarizing the main conclusions of this study:

a) Hot bottom burning in massive AGB stars coupled with *s*-process nucleosynthesis activated via the $^{22}$Ne neutron source and efficient third dredge-up allow the production of several short-lived nuclides that are found to be present in the early solar system. We can simultaneously match the observed abundances of $^{26}$Al, $^{41}$Ca, $^{60}$Fe, and $^{107}$Pd using a dilution factor of 1 part of AGB material per 300 parts of original solar nebula material, and a time interval between injection of the short-lived nuclides and formation of the first solid bodies in the solar system equal to 0.53 Myr.

b) Isotopic ratios of stable isotopes are only marginally modified by a massive AGB polluting source. The largest variations after dilution (2.4% at most) are predicted for the CNO isotopic ratios, and for the $^{46}Ca/^{40}Ca$ ratio.

c) There are no presolar stellar grains for which an origin in a massive AGB star has been confirmed. A possibility is that we still have not discovered grains from massive AGB stars because they may be much smaller than the grains currently analyzed in the laboratory (Bernatowicz et al. 2006, Nuth et al. 2006). The problem remains open.

In summary, taking the CAIs composition to be the protoplanetary disk isotopic abundances at the time of the earliest solid formation (see, e.g., Young et al. 2005), measured anomalies in refractory inclusions could be consistent with a picture where the first stages of the solar protoplanetary disk were enriched with the contribution of a nearby AGB star. Additional spectroscopic observations and nucleosynthesis modeling efforts for massive AGB stars must be pursued to disentangle the different stellar components contributing to the early composition of our solar system.

In this paper we have demonstrated that massive AGB stars can produce many of the SLN found in the solar system, in particular $^{26}$Al, $^{60}$Fe, and $^{41}$Ca, without the problem associated to SNII of overproducing $^{53}$Mn. We have also shown that, in certain mixing scenarios, AGB stars can produce these SLN in the right proportions. It is usually argued that it is unlikely that the SLN in the early solar system came from an AGB star because these stars are rarely found in star forming regions. We discuss this point in more detail in the appendix. We note in passing that some scenarios for the SNII origin of SLN are also facing a probability problem (Williams and Gaidos 2007, Gounelle and Meibom 2008). One may also consider that "as with any singular event [if a SN or an AGB star provided to the solar nebula the SLN], it is of little use to consider the *a priori* probability of this event" (Meyer and Zinner 2006). We should look at the evidence that primitive meteorites have preserved from the early solar system period, and provide tests for different scenarios regarding the production of SLNs. Further interdisciplinary studies are needed to address the fundamental problem of the birth of our solar system.

ACKNOWLEDGEMENTS

We thank Chris Matzner for invaluable help on the topic of star formation. We also thank Peter Anders, Soeren Larsen, Brad Gibson, Maurizio Busso, and Raquel Salmeron for discussion. We are very thankful to the referees, Ernst Zinner and Brad Meyer, for providing us very critical, but detailed and constructive reports. J.M.T.-R thanks *Consejo Superior de Investigaciones Científicas (CSIC)* for a JAE-Doc contract, and funding received from *Programa Nacional de Astronomía y Astrofísica* research project # ESP2007-61593. ML is supported by the *Netherland Organisation for Scientific Research (NWO)* via a VENI grant. AIK is supported by the *Australian Research Council* via an APD fellowship. We thank *NWO* and the *Netherlands Research School for Astronomy (NOVA)* for financial support for Amanda Karakas's visit to Utrecht.

APPENDIX: THE PLAUSIBILITY OF AGB POLLUTION

We start by citing the first page of the Cameron and Truran (1977) paper on "The Supernova Trigger for Formation of the Solar System": "We have found it to be a common reaction for people to ask: Is it not highly unlikely that a supernova should have gone off close to a region of formation of the solar system within a few million years of the time the event occurred?" The proposed solution to this problem was the idea that the formation of the solar system was triggered by the shock wave from a nearby supernova. The cause of its formation would then necessarily provide the presence of radioactive nuclei. The supernova trigger theory was based on two main premises: (1) the radioactive nuclei observed in the ESS could be made by a SNII, and

(2) supernova ejecta could trigger the collapse of a nearby molecular cloud, thus plausibly initiating the formation of many protostars. An AGB star can also satisfy the requirements needed to be considered a possible source for the SLN in the ESS in a trigger scenario: (1) nucleosynthesis in AGB stars can produce the needed abundances of SLN, as demonstrated by this work and that of Wasserburg and collaborators (Wasserburg et al. 1994, 2006), and (2) shock waves from AGB winds may be capable of triggering the collapse of the proto-solar cloud, as demonstrated by hydrodynamic simulations of a low-speed shock front (Boss 1995). This latter study, however, should be updated in the light of the resolution criteria proposed by Truelove et al. (1997). Also, this result does not necessarily confirm the AGB trigger scenario because these kinds of shocks are produced in star-forming regions by proto-stellar outflows. In any case, the idea of an AGB star trigger was favored for some time (Cameron 1984, 1993).

Subsequently, the AGB scenario was dismissed, primarily because Kastner and Myers (1994), using observational data, estimated that any giant molecular cloud (MC) located within 1 kpc of the Sun (from the list of Dame et al. 1987), has only about a 1% probability of encountering an AGB star in a 1 Myr period, which implies that AGB stars are relatively rare near star-forming regions today. As discussed by Kastner and Myers (1994), the probability of AGB-MC encounters would increase by an order of magnitude at the time of the formation of the solar system when using the bimodal initial mass function (IMF) proposed by Wyse and Silk (1987). However, galactic chemical evolution models no longer need this kind of IMF, because dual-infall models (e.g., Fenner and Gibson 2003) can generally account for observables (e.g., the so-called G-dwarf problem) that were once problematic. Kastner and Myers (1994) also state that "There is a significant (~70%) probability at the present epoch for a given cloud to be visited by an AGB star in ~$10^8$ yr". If this encounter "triggers multiple star formation, then AGB-induced proto-stars should exist in every molecular cloud" (Boss 1995). This probability would be lower when only considering massive AGB stars of ~4-8 $M_\odot$, however, a detailed re-analysis of this point is needed because the statistics of Kastner and Myers (1994) are very poor and completely dominated by only two stars. These authors concentrated on stars in the solar neighborhood, and noted that the extension to a larger volume in the Galactic disk would give a better estimation. A revised estimate (which is out of the scope of this work) should take into account the following:

1) The most massive AGB stars in our Galaxy pertain to the class of OH/IR stars, which are mainly concentrated in the Galactic disk (at low Galactic latitudes), as expected for a massive population (García-Hernández et al. 2007a and references therein), whereas lower-mass AGB stars display a higher scale height. Almost all massive AGB stars are found in the Galactic disk inside the solar circle and in the Galactic bulge, while there are fewer ones in the anti-center direction than in the direction of the Galactic center (Habing 1996). Hence, the extension of Kastner and Myers's work to a larger volume in the Galactic disk would include more of these stars and give a better estimation of the AGB-MC encounter probability.

2) The location where the Sun was born is still debated. It may have been born, for example, closer to the center of the Galaxy and then traveled to where it is now (see, e.g., Wielen et al. 1996). Thus, the AGB-MC encounter probability observed today in the solar neighborhood may not be really representative.

3) Only 28% of the known massive Galactic AGB stars (studied by García-Hernández et al. 2007a) are in the sample of Kastner and Myers. Their sample was based on the CO catalogs of Loup et al. (1993) and Jura and Kleinmann (1989), and not all known OH/IR stars are in these catalogs. Note that a conservative estimate of the total number of OH/IR stars (excluding the inner Galaxy where the crowding of stars is very high) is about ~2,000 (Habing 1996).

4) The luminosity selection criteria of Kastner and Myers may not be completely valid. They removed all stars in their sample with L<3,000 $L_{\odot}$ and L>60,000 $L_{\odot}$. However, it is well known that massive galactic AGB stars are strongly variable, with bolometric luminosity variations of several orders of magnitude (see, e.g., Engels et al. 1983). The mid- and far-infrared fluxes, which dominate the stellar luminosity of such extreme stars, can vary by more than 50% (e.g., García-Hernández et al. 2007b). This is confirmed by the available IRAS (the InfraRed Astronomical Satellite) measurements of extreme OH/IR stars from epoch to epoch, with only a few detailed infrared monitoring studies of some of these stars available in the literature (e.g., Engels et al. 1983). Thus, the luminosity variation of these stars is not well known. When considering the large uncertainty of their bolometric luminosities, together with the fact that Galactic distances are very uncertain, the number of massive AGB stars as candidates for encounters with MC may increase. Also, massive AGB stars can display important flux excess due to HBB (e.g., Whitelock et al., 2003) and they could be brighter than expected from theoretical predictions of AGB stars (e.g., Iben and Renzini 1983). For example, van Langevelde et al. (1990) measured luminosities between 4,300 and 97,000 $L_{\odot}$ in a small sample of massive AGB stars of our Galaxy.

5) At present, it is not known if luminous dusty obscured AGB stars - completely obscured at optical and near infrared (<3 microns) wavelengths - can be embedded in star forming regions of molecular clouds. For example, Spitzer telescope Galactic surveys (e.g., GLIMPSE) recently discovered a large number of faint stars with OH masers (an important fraction of them are expected to be massive AGB and post-AGB objects) that escaped detection by the IRAS mission (Engels 2007). Spitzer surveys of the Large Magellanic Cloud (SAGE; e.g., Meixner et al. 2006) and its surveys of other galaxies (SINGS) will be essential in order to elucidate the possible presence of these extreme objects embedded in dusty star forming regions.

Hence, the actual AGB-MC encounter probability may be higher than 1% per Myr. A future revision of this point according to the above mentioned points is needed.

# TABLES

Table 1. Radioactive isotopes detected in ESS materials and discussed in the present paper (adapted from Wasserburg et al. 2006).

| Parent | Half-Life (Myr) | reference ratio | ESS ratio |
|---|---|---|---|
| $^{10}$Be | 1.5 | $^{10}$Be/$^{9}$Be | $1\times10^{-3}$ |
| $^{26}$Al | 0.7 | $^{26}$Al/$^{27}$Al | $5\times10^{-5}$ |
| $^{36}$Cl | 0.3 | $^{36}$Cl/$^{35}$Cl | $5\times10^{-6}$ |
| $^{41}$Ca | 0.1 | $^{41}$Ca/$^{40}$Ca | $1.5\times10^{-8}$ |
| $^{53}$Mn | 3.7 | $^{53}$Mn/$^{55}$Mn | $6\times10^{-5}$; $5\times10^{-6}$ |
| $^{60}$Fe | 1.5 | $^{60}$Fe/$^{56}$Fe | $2\times10^{-6}$; $2\times10^{-7}$ |
| $^{107}$Pd | 6.5 | $^{107}$Pd/$^{108}$Pd | $2\times10^{-5}$ |
| $^{129}$I | 23 | $^{129}$I/$^{127}$I | $1\times10^{-4}$ |
| $^{182}$Hf | 13 | $^{182}$Hf/$^{180}$Hf | $2\times10^{-4}$ |

Table 2. Final envelope ratios for an AGB model of 6.5 $M_{\odot}$ and metallicity 0.02.

| Isotopic Ratio | $(N^R/N^I)_{ENV}$ |
|---|---|
| $^{26}Al/^{27}Al$ | $1.5\times10^{-2}$ |
| $^{41}Ca/^{40}Ca$ | $1.6\times10^{-4}$ |
| $^{60}Fe/^{56}Fe$ | $1.0\times10^{-3}$ |
| $^{107}Pd/^{108}Pd$ | $1.2\times10^{-2}$ |

Table 3. Ratios of the SLN considered here at different times as compared to those inferred from measurements of solar system samples. Numbers in brackets are imposed in order to derive $\Delta_1$ and $\Delta_2$.

| Ratio | no time interval | at $\Delta_1$ | at $\Delta_2$ | observed ESS |
|---|---|---|---|---|
| $^{26}Al/^{27}Al$ | $5.4\times10^{-5}$ | $3.2\times10^{-5}$ | $9.8\times10^{-8}$ | $5\times10^{-5}$ |
| $^{41}Ca/^{40}Ca$ | $5.2\times10^{-7}$ | $(1.5\times10^{-8})$ | - | $1.5\times10^{-8}$ |
| $^{60}Fe/^{56}Fe$ | $3.3\times10^{-6}$ | $2.6\times10^{-6}$ | $1.7\times10^{-7}$ | $2\times10^{-7}$ to $2\times10^{-6}$ |
| $^{107}Pd/^{108}Pd$ | $4\times10^{-5}$ | $3.8\times10^{-5}$ | $(2\times10^{-5})$ | $2\times10^{-5}$ |

Table 4. Predictions for anomalies for all stable isotopic ratios included in our network, in form of variations with respect to solar in parts per ten thousand ($\varepsilon$), together with their final AGB ratios ($Y^i_{AGB}/Y^j_{AGB}$), the AGB ratios diluted with solar system materials, and the solar ratios ($Y^i_{\odot}/Y^j_{\odot}$) we have used as reference and as initial in our calculations (Anders and Grevesse 1989).

| | $Y^i_{AGB}/Y^j_{AGB}$ | After dilution (f=3.3d-3) | $Y^i_{\odot}/Y^j_{\odot}$ | $\varepsilon^*$ |
|---|---|---|---|---|
| $^{13}C/^{12}C$ | $1.0276\times10^{-1}$ | $1.1292\times10^{-2}$ | $1.1112\times10^{-2}$ | 162 |
| $^{15}N/^{14}N$ | $3.2599\times10^{-5}$ | $3.5976\times10^{-3}$ | $3.6855\times10^{-3}$ | -239 |
| $^{17}O/^{16}O$ | $2.2891\times10^{-3}$ | $3.8644\times10^{-4}$ | $3.8132\times10^{-4}$ | 134 |
| $^{18}O/^{16}O$ | $1.7230\times10^{-6}$ | $2.0030\times10^{-3}$ | $2.0083\times10^{-3}$ | -27 |
| $^{21}Ne/^{20}Ne$ | $1.3520\times10^{-4}$ | $2.4198\times10^{-3}$ | $2.4274\times10^{-3}$ | -31 |
| $^{22}Ne/^{20}Ne$ | $8.0744\times10^{-2}$ | $7.3132\times10^{-2}$ | $7.3107\times10^{-2}$ | 3.5 |
| $^{25}Mg/^{24}Mg$ | $2.1223\times10^{-1}$ | $1.2644\times10^{-1}$ | $1.2615\times10^{-1}$ | 22 |
| $^{26}Mg/^{24}Mg$ | $2.7860\times10^{-1}$ | $1.3962\times10^{-1}$ | $1.3916\times10^{-1}$ | 33 |
| $+^{26}Al$** | | | | 33 |
| $^{29}Si/^{28}Si$ | $5.3335\times10^{-2}$ | $5.0652\times10^{-2}$ | $5.0643\times10^{-2}$ | 1.8 |
| $^{30}Si/^{28}Si$ | $3.6519\times10^{-2}$ | $3.3629\times10^{-2}$ | $3.3619\times10^{-2}$ | 2.9 |
| $^{33}S/^{32}S$ | $8.1936\times10^{-3}$ | $7.8943\times10^{-3}$ | $7.8933\times10^{-3}$ | 1.3 |
| $^{36}S/^{32}S$ | $2.7429\times10^{-4}$ | $2.1080\times10^{-4}$ | $2.1059\times10^{-4}$ | 10 |
| $^{35}Cl/^{37}Cl$ | 2.2559 | 3.1289 | 3.1330 | -13 |
| $^{36}Ar/^{40}Ar$ | $1.4149\times10^{3}$ | $3.3851\times10^{3}$ | $3.4008\times10^{3}$ | -46 |
| $^{40}K/^{39}K$ | $3.5329\times10^{-3}$ | $1.5648\times10^{-3}$ | $1.5582\times10^{-3}$ | 43 |
| $^{41}K/^{39}K$ | $7.9827\times10^{-2}$ | $7.2168\times10^{-2}$ | $7.2142\times10^{-2}$ | 3.6 |
| $+^{41}Ca$** | | | | 7.1 |
| $^{42}Ca/^{40}Ca$ | $7.2129\times10^{-3}$ | $6.6707\times10^{-3}$ | $6.6689\times10^{-3}$ | 2.7 |
| $^{43}Ca/^{40}Ca$ | $1.5433\times10^{-3}$ | $1.3937\times10^{-3}$ | $1.3932\times10^{-3}$ | 3.6 |
| $^{46}Ca/^{40}Ca$ | $1.7236\times10^{-4}$ | $4.0970\times10^{-5}$ | $4.0534\times10^{-5}$ | 107 |
| $^{49}Ti/^{48}Ti$ | $8.0273\times10^{-2}$ | $7.4567\times10^{-2}$ | $7.4548\times10^{-2}$ | 2.5 |
| $^{50}Ti/^{48}Ti$ | $8.4062\times10^{-2}$ | $7.3444\times10^{-2}$ | $7.3409\times10^{-2}$ | 4.8 |
| $^{54}Cr/^{52}Cr$ | $3.3922\times10^{-2}$ | $2.8212\times10^{-2}$ | $2.8194\times10^{-2}$ | 6.7 |
| $^{57}Fe/^{56}Fe$ | $2.5517\times10^{-2}$ | $2.4006\times10^{-2}$ | $2.4001\times10^{-2}$ | 2.1 |
| $^{58}Fe/^{56}Fe$ | $6.9115\times10^{-3}$ | $3.0669\times10^{-3}$ | $3.0541\times10^{-3}$ | 42 |
| $^{60}Ni/^{58}Ni$ | $4.0466\times10^{-1}$ | $3.8287\times10^{-1}$ | $3.8280\times10^{-1}$ | 1.9 |
| $+^{60}Fe$** | | | | 4.9 |
| $^{61}Ni/^{58}Ni$ | $2.4055\times10^{-2}$ | $1.6555\times10^{-2}$ | $1.6530\times10^{-2}$ | 15 |
| $^{62}Ni/^{58}Ni$ | $7.5474\times10^{-2}$ | $5.2598\times10^{-2}$ | $5.2522\times10^{-2}$ | 14 |
| $^{64}Ni/^{58}Ni$ | $2.3605\times10^{-2}$ | $1.3361\times10^{-2}$ | $1.3327\times10^{-2}$ | 25 |
| $^{65}Cu/^{63}Cu$ | $3.7691\times10^{-1}$ | $4.4554\times10^{-1}$ | $4.4590\times10^{-1}$ | -8.2 |
| $^{66}Zn/^{64}Zn$ | $6.6076\times10^{-1}$ | $5.7443\times10^{-1}$ | $5.7415\times10^{-1}$ | 5.0 |
| $^{67}Zn/^{64}Zn$ | $1.0288\times10^{-1}$ | $8.4394\times10^{-2}$ | $8.4333\times10^{-2}$ | 7.2 |
| $^{68}Zn/^{64}Zn$ | $4.8486\times10^{-1}$ | $3.8518\times10^{-1}$ | $3.8485\times10^{-1}$ | 8.5 |
| $^{70}Ge/^{74}Ge$ | $6.4369\times10^{-1}$ | $5.6236\times10^{-1}$ | $5.6202\times10^{-1}$ | 6.1 |
| $^{73}Ge/^{74}Ge$ | $2.1916\times10^{-1}$ | $2.1387\times10^{-1}$ | $2.1384\times10^{-1}$ | 1.0 |
| $^{76}Ge/^{74}Ge$ | $1.7057\times10^{-1}$ | $2.1357\times10^{-1}$ | $2.1375\times10^{-1}$ | -8.5 |
| $^{74}Se/^{80}Se$ | $1.3431\times10^{-2}$ | $1.7791\times10^{-2}$ | $1.7810\times10^{-2}$ | -10 |
| $^{76}Se/^{80}Se$ | $2.2714\times10^{-1}$ | $1.8149\times10^{-1}$ | $1.8129\times10^{-1}$ | 11 |
| $^{82}Se/^{80}Se$ | $1.4455\times10^{-1}$ | $1.8419\times10^{-1}$ | $1.8437\times10^{-1}$ | -9.3 |
| $^{81}Br/^{79}Br$ | 1.1236 | $9.7342\times10^{-1}$ | $9.7290\times10^{-1}$ | 5.4 |
| $+^{81}Kr$** | | | | 5.4 |
| $^{78}Kr/^{84}Kr$ | $4.5996\times10^{-3}$ | $5.9488\times10^{-3}$ | $5.9545\times10^{-3}$ | -9.6 |
| $^{80}Kr/^{84}Kr$ | $3.2680\times10^{-2}$ | $3.8845\times10^{-2}$ | $3.8871\times10^{-2}$ | -6.7 |
| $^{82}Kr/^{84}Kr$ | $2.6962\times10^{-1}$ | $2.0057\times10^{-1}$ | $2.0028\times10^{-1}$ | 15 |
| $^{83}Kr/^{84}Kr$ | $1.9533\times10^{-1}$ | $2.0072\times10^{-1}$ | $2.0074\times10^{-1}$ | -1.1 |
| $^{86}Kr/^{84}Kr$ | $4.7475\times10^{-1}$ | $3.0574\times10^{-1}$ | $3.0503\times10^{-1}$ | 23 |
| $^{87}Rb/^{85}Rb$ | $7.3056\times10^{-1}$ | $4.1370\times10^{-1}$ | $4.1212\times10^{-1}$ | 38 |
| $^{87}Sr/^{86}Sr$ | $5.6247\times10^{-1}$ | $6.5044\times10^{-1}$ | $6.5088\times10^{-1}$ | -6.7 |
| $^{87}Rb/^{86}Sr$ | 1.6131 | $9.1284\times10^{-1}$ | $9.0935\times10^{-1}$ | 38 |
| $^{88}Sr/^{86}Sr$ | 7.0721 | 8.3596 | 8.3660 | -7.7 |
| $^{91}Zr/^{90}Zr$ | $2.2515\times10^{-1}$ | $2.1815\times10^{-1}$ | $2.1813\times10^{-1}$ | 1.2 |
| $^{92}Zr/^{90}Zr$ | $3.4699\times10^{-1}$ | $3.3414\times10^{-1}$ | $3.3410\times10^{-1}$ | 1.4 |
| $^{96}Zr/^{90}Zr$ | $8.2476\times10^{-2}$ | $5.4650\times10^{-2}$ | $5.4549\times10^{-2}$ | 19 |
| $^{96}Mo/^{98}Mo$ | $7.1668\times10^{-1}$ | $6.9097\times10^{-1}$ | $6.9087\times10^{-1}$ | 1.3 |
| $^{100}Mo/^{98}Mo$ | $3.7220\times10^{-1}$ | $3.9984\times10^{-1}$ | $3.9994\times10^{-1}$ | -2.5 |
| $^{96}Ru/^{102}Ru$ | $1.6783\times10^{-1}$ | $1.7516\times10^{-1}$ | $1.7518\times10^{-1}$ | -1.4 |
| $^{98}Ru^{102}Ru$ | $5.7102\times10^{-2}$ | $5.9528\times10^{-2}$ | $5.9536\times10^{-2}$ | -1.4 |
| $^{99}Ru/^{102}Ru$ | $3.8532\times10^{-1}$ | $4.0142\times10^{-1}$ | $4.0148\times10^{-1}$ | -1.4 |
| $+^{99}Tc$** | | | | 13 |
| $^{100}Ru/^{102}Ru$ | $4.2124\times10^{-1}$ | $3.9813\times10^{-1}$ | $3.9805\times10^{-1}$ | 2.0 |
| $^{101}Ru/^{102}Ru$ | $5.1596\times10^{-1}$ | $5.3745\times10^{-1}$ | $5.3753\times10^{-1}$ | -1.4 |
| $^{102}Pd^{106}Pd$ | $3.5479\times10^{-2}$ | $3.7360\times10^{-2}$ | $3.7366\times10^{-2}$ | -1.7 |
| $^{104}Pd/^{106}Pd$ | $4.2531\times10^{-1}$ | $4.0794\times10^{-1}$ | $4.0789\times10^{-1}$ | 1.4 |
| $^{105}Pd/^{106}Pd$ | $7.8222\times10^{-1}$ | $8.1555\times10^{-1}$ | $8.1567\times10^{-1}$ | -1.4 |
| $^{110}Pd/^{106}Pd$ | $4.4357\times10^{-1}$ | $4.2895\times10^{-1}$ | $4.2890\times10^{-1}$ | 1.2 |

Footnotes:

* Variation with respect to solar per ten thousand. Only ratios for which $\varepsilon > 1$ are listed.

** In these rows the abundance of the short-lived nucleus is added to that of the daughter stable nucleus to calculate the $\varepsilon$ value.

Table 5. Model predictions for other SLN of interest.

| | AGB ratio | no time interval | at $\Delta_1$ | at $\Delta_2$ |
|---|---|---|---|---|
| $^{81}Kr/^{82}Kr$ | $3.4\times10^{-4}$ | $2.0\times10^{-6}$ | $3.9\times10^{-7}$ | $5.5\times10^{-15}$ |
| $^{93}Zr/^{92}Zr$ | $4.4\times10^{-2}$ | $1.6\times10^{-4}$ | $1.3\times10^{-4}$ | $8.6\times10^{-6}$ |
| $^{99}Tc/^{100}Ru$ | $2.4\times10^{-2}$ | $8.5\times10^{-5}$ | $1.4\times10^{-5}$ | $3.3\times10^{-14}$ |

# FIGURES

Figure 1. Evolution of the SLN isotopic ratios at the surface of our 6.5 $M_\odot$ Z=0.02 model star on a logarithmic scale as function of the TP number.

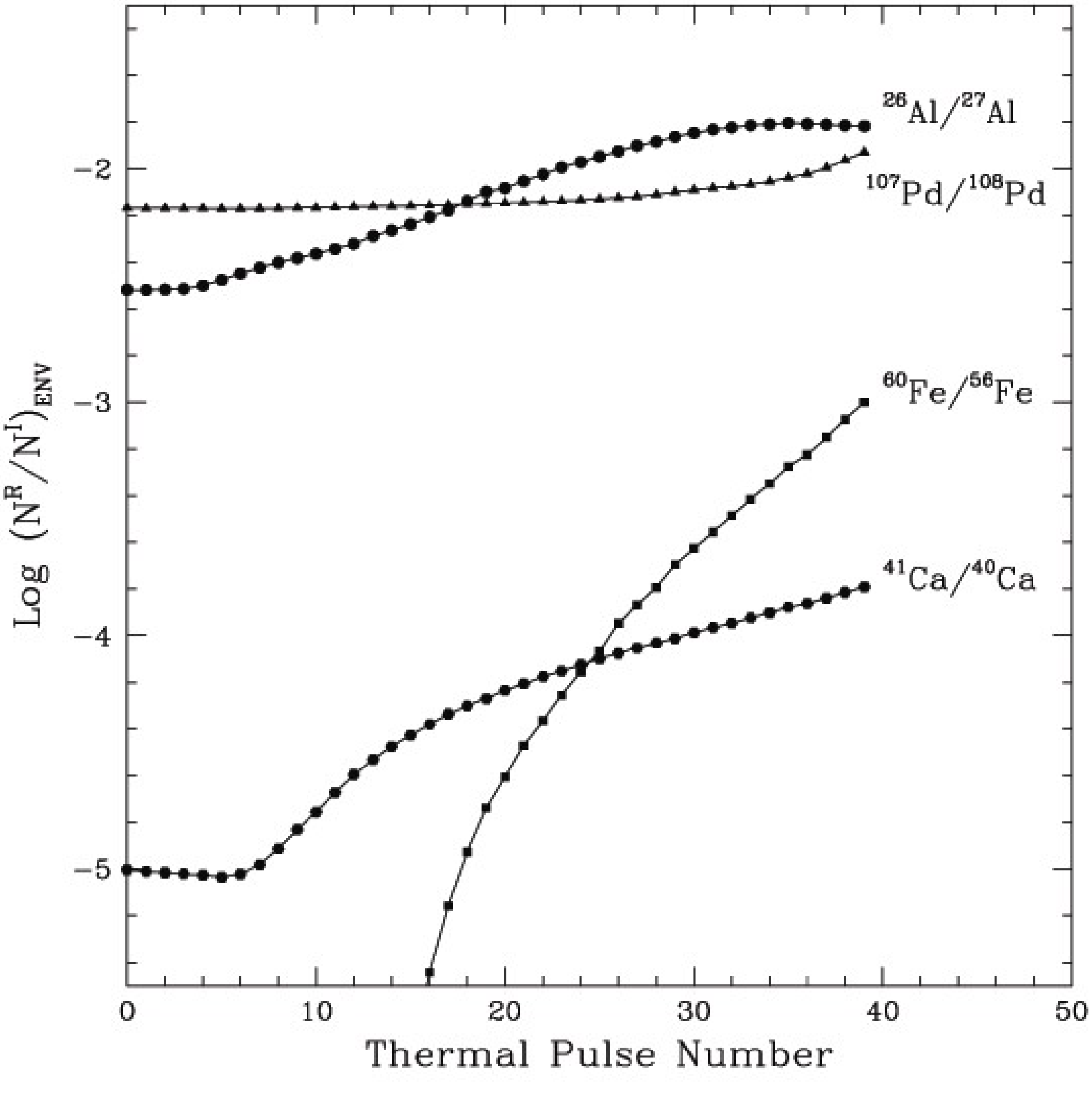